# Enhanced metallic properties of SrRuO$_3$ thin films via kinetically controlled pulsed laser epitaxy

J. Thompson[1], J. Nichols[2], S. Lee[2], S. Ryee[3], J. H. Gruenewald[1], J. G. Connell[1], M. Souri[1], J. M. Johnson[4], J. Hwang[4], M. J. Han[3], H. N. Lee[2], D.-W. Kim[5], and S. S. A. Seo[1]

[1]Department of Physics and Astronomy, University of Kentucky, Lexington, KY 40506, USA

[2]Materials Science and Technology Division, Oak Ridge National Laboratory, Oak Ridge, TN 37831, USA

[3]Department of Physics, Korea Advanced Institute of Science and Technology (KAIST), Daejeon 305-701, Korea

[4]Department of Materials Science and Engineering, The Ohio State University, Columbus, OH 43212, USA

[5]Department of Physics, Ewha Womans University, Seoul 120-750, Korea

## Abstract

Metal electrodes are a universal element of all electronic devices.  Conducting SrRuO$_3$ (SRO) epitaxial thin films have been extensively used as electrodes in complex-oxide heterostructures due to good lattice mismatches with perovskite substrates.  However, when compared to SRO single crystals, SRO thin films have shown reduced conductivity and Curie temperatures ($T_C$), which can lead to higher Joule heating and energy loss in the devices.  Here, we report that high-quality SRO thin films can be synthesized by controlling the plume dynamics and growth rate of pulsed laser epitaxy (PLE) with real-time optical spectroscopic monitoring.  The SRO thin films grown under the kinetically controlled conditions, down to ca. 16 nm in thickness, exhibit both enhanced conductivity and $T_C$ as compared to bulk values, due to their improved stoichiometry and a strain-mediated increase of the bandwidth of Ru 4$d$ electrons.  This result provides a direction for enhancing the physical properties of PLE-grown thin films and paves a way for improved device applications.

Complex-oxide materials with intriguing properties have dawned a new era of the so-called 'all epitaxial functional devices' such as non-volatile memories[1-4], high speed switching devices[5-7], piezoelectric nano-generators[8], and ultraviolet lasers[9]. The recent advances in epitaxial growth techniques for complex oxides enabled investigations of high quality ultrathin films with thickness as thin as a few nanometers, yielding novel physical properties[10]. Exploration of the device applicability requires suitable metal electrodes that maintain such emergent physical properties. Normal metals and alloys such as Pt, Au, Ag, and Cu with high electrical conductivity lack interface adhesion and structural compatibility with complex oxides, which is essential for the fabrication of high performance devices. $SrRuO_3$ (SRO) is one of the most extensively studied and widely used metallic oxides[11,12]. The perovskite structure yields SRO to have excellent chemical stability, which makes it an ideal electrode for oxide heterostructures[13]. However, so far, SRO thin films have shown inferior metallic properties as compared to their bulk counterparts. SRO single crystals typically have a Curie temperature ($T_C$) around 160-165 K and a room temperature resistivity of ~150 μΩcm (Refs. 14-17), yet no films have been synthesized which maintain these properties (see Table 1). For example, SRO thin films grown on $GdScO_3$ (GSO) substrates have a significantly higher resistivity of ~650 μΩcm at room temperature and low $T_C$ (100-130 K) (Refs. 18,19). SRO thin films grown on $SrTiO_3$ (STO) substrates also exhibit a high room-temperature resistivity of ~225 μΩcm and low $T_C$, which approaches 150 K for thickness above ~25 nm (Refs. 20-23). Hence, it is essential to ask, "Are the transport properties of SRO thin films inherently inferior to SRO single crystals, or is there a way to enhance the metallic properties in the thin film limit?"

To answer these questions, we have investigated epitaxial thin films of SRO with various thicknesses (1-65 nm) grown on atomically flat GSO $(110)_o$ substrates (see Fig. S1 of supplementary material), using pulsed laser epitaxy (PLE). The pseudo-cubic lattice parameters of SRO and GSO are 3.93 Å and 3.96 Å, respectively. Therefore, the SRO thin films grown on GSO substrates will experience in-plane tensile strain of +0.76 % (see Fig. S2 of supplementary material). According to our first-principles calculations[24,25] (see Fig. S3 of supplementary material), SRO thin films under tensile strain are expected to exhibit enhanced metallic properties as a result of the increase in the Ru 4$d$ electron bandwidth and average exchange energy ($J_{avg.}$). Hence, we have chosen GSO substrates to study the effect of kinetically controlled growth on the physical properties of SRO thin films. We have prepared atomically flat GSO (110)$_o$ substrates (from CrysTec GmbH) by annealing at 1000°C in air for 1hr.[26] The SRO thin films are deposited at 600°C in an oxygen partial pressure of 100 mTorr, with a KrF excimer laser ($\lambda$ = 248 nm) with a fluence of 1.6 J/cm$^2$ at 10 Hz using a ruthenium rich polycrystalline target.

In order to control the deposition rate, we have used a variety of laser spot sizes (0.16-0.41 mm$^2$) by changing the aperture size in our laser optics. Note that the shape of the laser spot (square) was the same for all spot sizes. In general, a larger (smaller) laser spot size produces a larger (smaller) PLE plume; therefore by changing the size of the laser spot, we can effectively control the deposition rate (for technical details, see Ref. 27). Note that the deposition rates we have used are between 150-800 pulses/u.c. (0.027-0.005nm/sec), which are significantly slower than the typical deposition rates (10-50 pulses/u.c.) of conventional pulsed laser deposition. Figure 1 shows that the deposition rate can be controlled by keeping our growth parameters fixed and only changing the laser spot size. Using an isotropic slab model[28] with the complex dielectric functions of a SRO thin film and a GSO substrate, as shown in Figs. 1 (a) and 1 (b), respectively, the real-time thickness of SRO thin films was monitored using *in-situ* optical spectroscopic ellipsometry[28] and the deposition rate was determined for each laser spot size. Figure 1 (c) shows the SRO *in-situ* film thickness as a function of time with the laser spot sizes of 0.16 mm$^2$, 0.30 mm$^2$, and 0.41 mm$^2$. The total thickness of the SRO thin films was also confirmed from the interference fringes in the x-ray diffraction (XRD) $\theta$-$2\theta$ scans. Figure 1 (d) shows the room temperature optical conductivity spectrum of a SRO thin film, which is consistent with a reference spectrum[29].

SRO thin films display a change in the crystal structure above 16 nm. Figure 2 (a) shows the XRD *$\theta$-$2\theta$* scans for our SRO thin films, which reveal the out-of-plane $(hh0)_o$ reflections of the orthorhombic phase. The inset of Fig. 2 (a) shows a rocking curve of a 16 nm-thick SRO thin film, which has a full width at half maximum (FWHM) value of 0.06 °, indicating good crystallinity of the film. We performed the XRD reciprocal space mapping (RSM) around the GSO 620 and 260 reflections, and define $Q_{//}$ along the in-plane $[1\text{-}10]_o$ direction, and $Q_\perp$ along the out-of-plane $[110]_o$ direction (see Fig. S4 of supplementary material). For all of our films, the 620 and 260 film peaks are at the same $Q_{//}$ position as the substrate, indicating they are coherently strained, whereas the position of the 260 peak moves to larger $Q_\perp$ with decreasing film thickness. Kan, *et al.* (Ref. 18) have reported a similar result for their SRO films grown on GSO substrates. Our RSM data indicate a structural change from an orthorhombic (monoclinic) to a tetragonal phase, as the SRO thin film thickness is increased above 16 nm. Cross-sectional scanning transmission electron microscopic images of a 65 nm-thick SRO thin film are shown with the beam along the $[1\text{-}10]_o$ (Fig. 2 (b)) and the $[001]_o$ (Fig. 2 (c)) directions. A sharp interface (red arrows) is observed between the GSO substrate and SRO thin film, and there are no indications of misfit dislocations or defects.

The resistivity and $T_C$ of thicker SRO thin films are similar to SRO bulk single crystals. The dc-transport behavior of SRO thin films is shown in Figure 3 as a function of temperature ($\rho(T)$). While a 1 nm thick film is insulating, SRO thin films with increased thickness show a clear metallic behavior. Note that $\rho(T)$ is significantly reduced above ca. 16 nm and it is very similar to that of bulk crystals[17]. Moreover, as summarized in Table I, the room temperature resistivity ($\rho(300K)$) of our SRO thin films are smaller than any previous reports of SRO single crystals and thin films. The seemingly surprising improvement over the single crystal resistivity is in agreement with our first-principles calculations, which suggest an enhanced conductivity of SRO under tensile strain (see Fig. S3 of supplementary material). SRO is an itinerant ferromagnet described by the Ru 4$d$ conduction bands using the Stoner model. The "kink" visible in the $\rho(T)$ data is due to the suppression of spin scattering as SRO transitions to a ferromagnetic state and represents the $T_C$ of SRO thin films. The $T_C$ is estimated by taking the first derivative of the resistivity ($d\rho/dT$), as shown in the inset of Figs. 3, S5, S6 (supplementary material). The $T_C$ values of our SRO thin films are close to (or higher than) the previously reported values of SRO single crystals (Refs. 14-16) and compressive-strained SRO thin films (Ref. 20-22).

The $T_C$ values gradually increase as the thickness is increased, reaching a maximum (~ 163 K) at 16 nm, and remains approximately constant above this thickness. Figure 4 (a) shows the estimated $T_C$'s as a function of thickness. Note that these $T_C$ values are significantly higher than previously reported $T_C$'s for SRO thin films under compressive (open triangles)[20] and tensile (open squares) strain[18], as shown in Fig. 4(a) for comparison. Although the room temperature resistivity and $T_C$ of our films are enhanced from other SRO thin films, the residual resistivity ratio (RRR) remains low (see Table I and Fig. S7 (supplementary material)). This is likely due to the appearance of a resistivity up-turn at low temperatures (< 20K) in SRO thin films, which has been observed in previous studies, but its origin is not fully understood at this moment.

To investigate the discrepancy in the enhanced metallic properties observed in our SRO thin films compared to that of other thin film reports, we have performed a test growth of a 15 nm-thick SRO thin film with an increased laser fluence of ~3.0 J/cm$^2$. Note that the optimal SRO thin films are grown at ~1.6 J/cm$^2$. The XRD $\theta$-$2\theta$ scan of the test-grown SRO thin film shows a slight increase in the out-of-plane lattice parameter (decrease in $2\theta$ values) compared to an optimal 16 nm-thick SRO thin film, as shown in Fig. 4 (b), while the FWHM of the two thin films rocking curves are very similar, indicating that they both are homogeneous with comparable crystalline quality. Surprisingly, the $\rho(T)$ data show that the 15 nm-thick SRO thin film grown at the high fluence (~3.0 J/cm$^2$) has roughly doubled the resistivity, as shown in Fig. 4 (c), and a significantly lower $T_C$ (~ 133 K) than those from the 16 nm-thick SRO thin film.

It is well known that Ru vacancies can be formed in SRO thin films due to the volatile nature of the Ru atom. According to Dabrowski et al.[30], Ru vacancies have a profound impact on the metallic and magnetic properties of SRO thin films, i.e. decreasing $T_C$ down to 86 K for only a 6 % change in Ru site occupancy. Therefore, we consider that the laser fluence in PLE may affect the stoichiometry of SRO thin films. Unfortunately, because the change in stoichiometry is so small, microscopic characterization measurements are not capable of resolving the differences in our films. However, the transport and magnetic properties of SRO thin films can be significantly affected as shown above. It is also known that structural distortions and symmetry mismatch across interface boundaries, and other interfacial effects can deteriorate electrical properties of complex-oxide thin films and heterostructures[31-33]. However, in our SRO thin films grown on GSO substrates, the laser beam spot size and, hence, the film deposition rate are shown to have a more significant role in their transport properties than the interfacial contributions (see Fig. S7 of supplementary material).

In order to verify the effects of the kinetically controlled growth rate on the properties of SRO thin films, we have grown two samples of similar thickness but with two different laser spot sizes. A 5 nm-thick SRO thin film was grown using the smallest laser spot size (0.16 mm$^2$) with an extremely slow growth rate (~800 pulses/u.c.). This film is compared to a 6 nm-thick film which was grown using a laser spot size of 0.35 mm$^2$ resulting in ~270 pulses/u.c. Our dc-transport measurements show that the resistivity of the 5 nm film is reduced compared to the 6 nm-thick film (Fig. S8 of supplementary material). This is surprising because for our films grown with the same laser spot size (0.35 mm$^2$) [13], the resistivity increases as the film thickness decreases. Hence, the moderated PLE techniques, i.e. controlled growth rate and laser spot size, are the key for the enhanced transport properties of SRO thin films. Lee *et al.*[27] has shown that the laser spot size plays a significant role in the oxygen stoichiometry of STO thin films fabricated with PLE, and our result shows that the laser spot size can also have an impact on the cation stoichiometry (i.e. Ru ions). Although it is an arduous task to precisely measure the stoichiometric ratio of thin films directly, by comparing data from samples intentionally grown with Ru or O vacancies[30,34], we have shown that the kinetically controlled (laser spot size) PLE deposition is a fundamental component of fabricating the highest quality complex oxide films and heterostructures.

This work dismisses the notion that SRO thin films are inherently inferior to single crystals, which has become a generally accepted problem in the solid state community. We have discovered that it is possible to achieve SRO thin films with metallic properties similar to SRO single crystals by adjusting the laser spot size and effectively reducing the laser plume kinetic energy, thereby improving the overall stoichiometry of the films. Our results show that by controlling the laser spot size (laser plume energy), it is possible to fabricate epitaxial thin film electrodes for functional oxide devices, which do not hinder the functionality of the device as a result of degraded metallic properties.

See supplementary material for the results of atomic force microscopy, first-principles calculations, XRD RSM, and additional transport measurements.

**Acknowledgements**

We acknowledge the support of National Science Foundation grant DMR-1454200 for the sample syntheses and characterizations. S.R. and M.J.H. were supported by Basic Science Research Program through NRF (2014R1A1A2057202) and the computing resource is supported by KISTI (KSC-2014-C2-046). The XRD measurements performed at ORNL were supported by the U.S. Department of Energy, Office of Science, Basic Energy Sciences, Materials Science and Engineering Division.

|  | $\rho$ (2 K) ($\mu\Omega$cm) | $\rho$ (300 K) ($\mu\Omega$cm) | RRR | $T_C$ (K) |
|---|---|---|---|---|
| This report, SRO/GSO (tensile strain), $t > 15$ nm | 25-30 | 125-140 | 5-6 | 160-163 |
| SRO single crystals[14-17] | 1-15 | 150-200 | 20-192 | 160-165 |
| SRO/STO (comp. strain)[20,21] (1-50 u.c.) | 25-80 | 225-300 | 2-14 | 130-150 |
| SRO/GSO (tensile strain)[18] (27-64 u.c.) | 350-375 | 650-700 | 2-4 | 100-130 |

**Figure Captions**

**FIG 1.** Real-time monitoring of the thin film thickness via *in-situ* optical spectroscopic ellipsometry. Schematic of the sample geometry with the associated *in-situ* real (blue) and imaginary (red) dielectric functions as a function of photon energy for **(a)** SRO thin film and **(b)** GSO substrate, collected after and before deposition, respectively. **(c)** Real time thickness of a SRO thin film extracted from the *in-situ* spectroscopic ellipsometry data using a single slab model during the growth. The red arrows indicate the start and stop point for deposition and the red asterisks represent when the deposition is stopped to change the laser spot size. **(d)** Comparison of room-temperature optical conductivity spectrum of our SRO thin film to the data previously reported for SRO[29].

**FIG 2.** X-Ray Diffraction patterns and cross-sectional High Resolution Scanning Transmission Electron Microscopy images obtained for films of $SrRuO_3$ deposited on $GdScO_3$ (110) substrates. **(a)** Out-of-plane $\theta$-$2\theta$ XRD patterns for SRO films around the $(220)_o$ peak, of thickness ranging from 6-65 nm. The inset shows a typical rocking curve for all of the films in this thickness range. **(b,c)** High resolution Scanning Transmission Electron Microscopy images of cross-sections with the beam along the $(1-10)_o$ **(b),** and $(001)_o$ **(c),** directions.

**FIG 3.** *dc*-Transport data for $SrRuO_3$ films of various thicknesses. Resistivity, $\rho(T)$, as a function of temperature for films of 1, 6, 16 and 32 nm thickness. For comparison we have also included digitized data from our references, for SRO single crystals (Ref. 17). The inset shows the derivative of the resistivity as a function of temperature for our 16 nm sample as well as the digitized data for single crystals (Ref. 17) and a SRO/STO (compressive strain) film (Ref. 20). We can clearly see that the $T_C$ and resistivity of our 16 nm sample is similar to the values reported for single crystals, and significantly improved from SRO/STO.

**FIG 4.** Effects of moderated deposition rate on the Curie temperature ($T_C$) and physical properties. **(a)** The Curie temperature versus the film thickness is shown as red filled circles. The open triangles and squares represent previously reported values for SRO/STO (Ref. 20) and SRO/GSO (Ref. 18) films (compressive and tensile strain), respectively. The $T_C$ reaches a maximum of ~163 K at 16 nm and remains essentially constant above this thickness. This value is significantly higher than previous reports for SRO films under compressive or tensile strain, and is even comparable to the values for single crystals. The green shaded area highlights the thickness region where the conductivity and $T_C$ become maximum and comparable to single crystals (black dashed lines). **(b)** Comparison of the XRD $\theta$-$2\theta$ data of two films grown at 350 pulses per unit cell but with different laser fluence. The red curve shows the 16 nm film (~1.6 J/cm$^2$) and the black curve shows the 15 nm film (~3 J/cm$^2$). **(c)** Comparison of the $\rho(T)$ data for the same two samples.

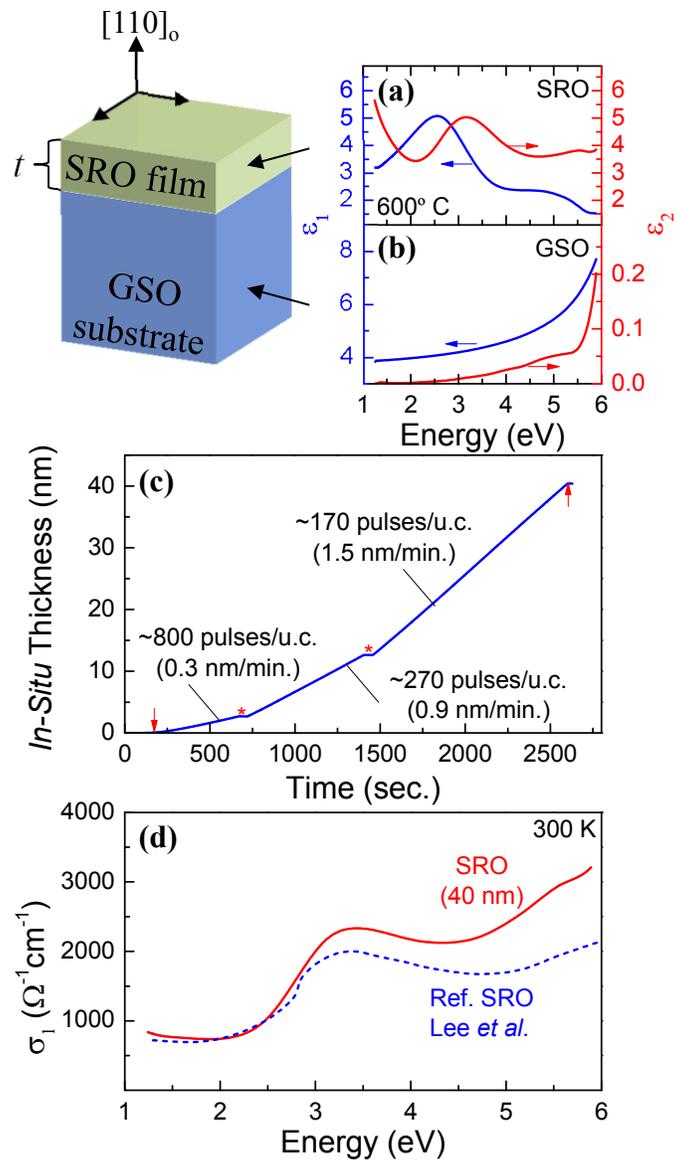

Figure 1
J. Thompson et al.

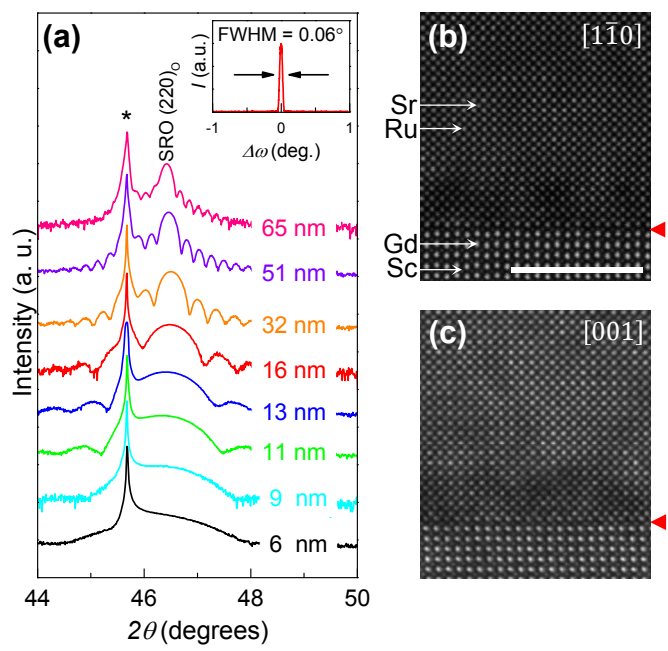

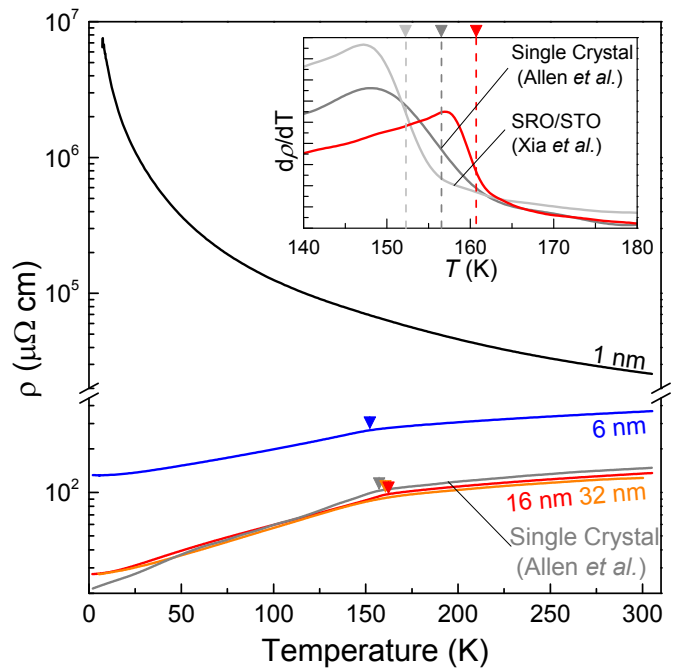

Figure 3
J. Thompson *et al.*

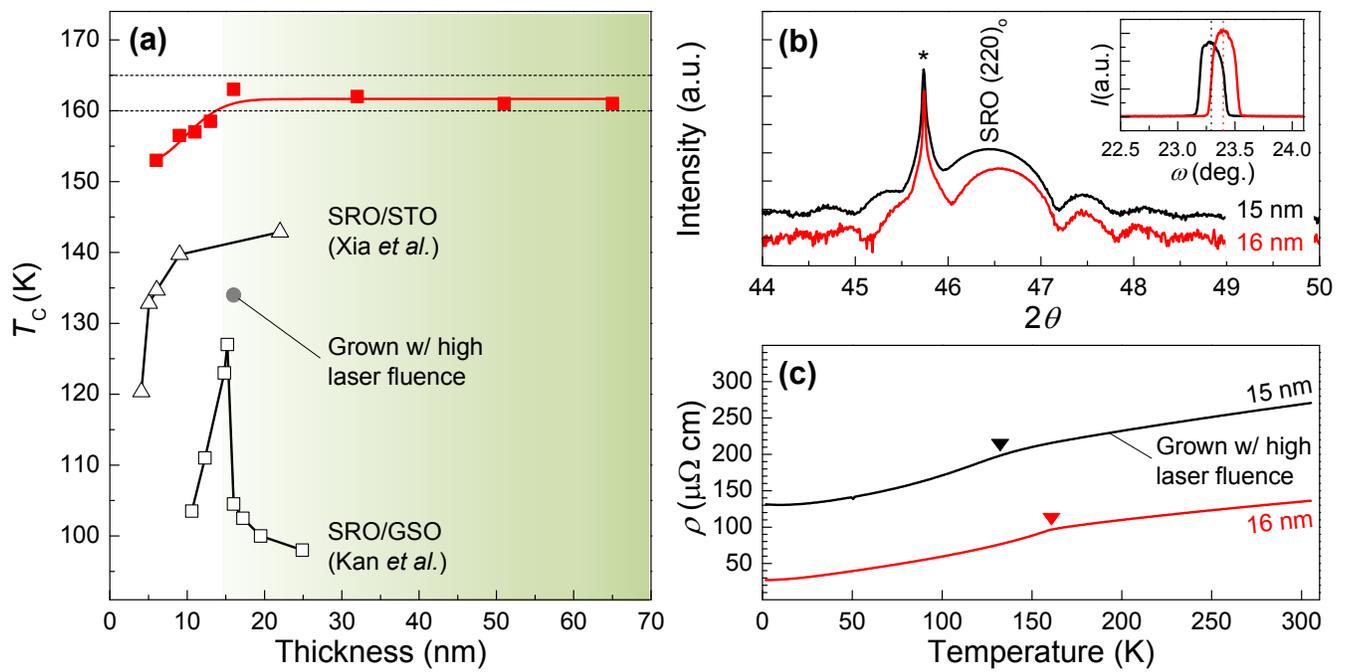

Figure 4
J. Thompson *et al.*

# Supplementary Material

**Enhanced metallic properties of SrRuO$_3$ thin films via kinetically controlled pulsed laser epitaxy**

J. Thompson *et al.*

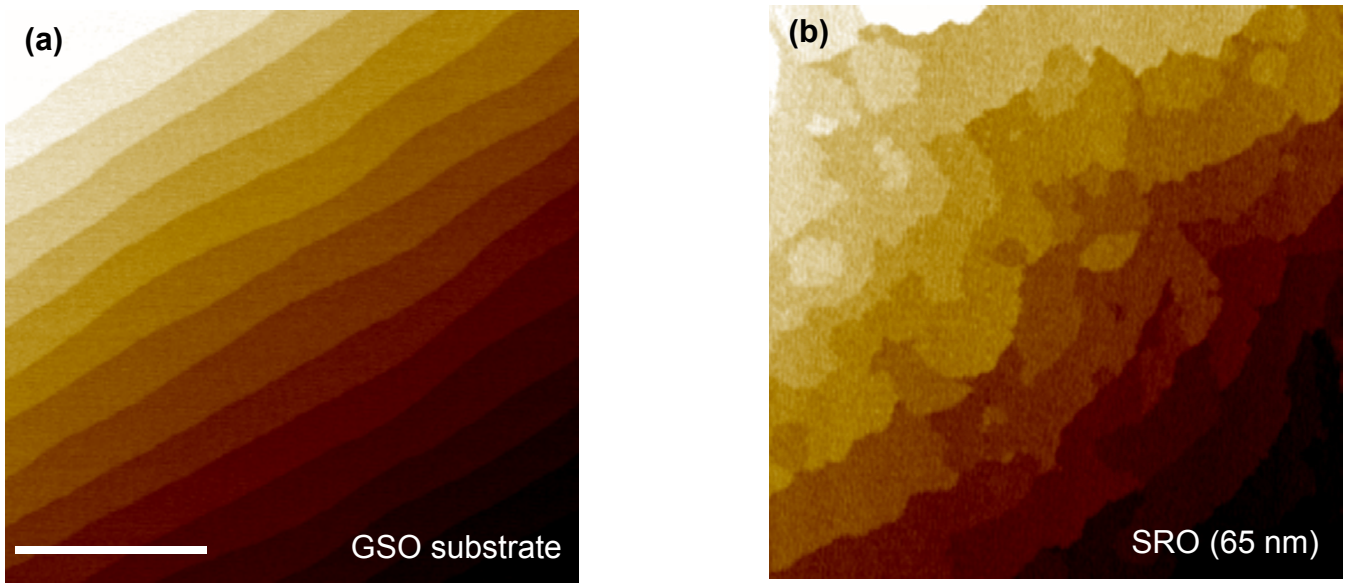

**Fig. S1. AFM Topography (3×3 μm) of GSO substrate and corresponding SRO film (65 nm)** **(a)** GSO substrate after annealing in air for 1 hour at 1000˚C. Our GSO substrates are atomically flat, having a typical surface roughness, $R_q$ = 1.4 nm, unit cell step height and miscut angle less than 0.1˚. **(b)** 65 nm SRO epi-thin film imaged immediately after deposition. The films all had similar surface roughness (Rq ≈ 1.6 nm) which is comparable to the substrate. The white line in **(a)** is a scale bar representing 1 μm.

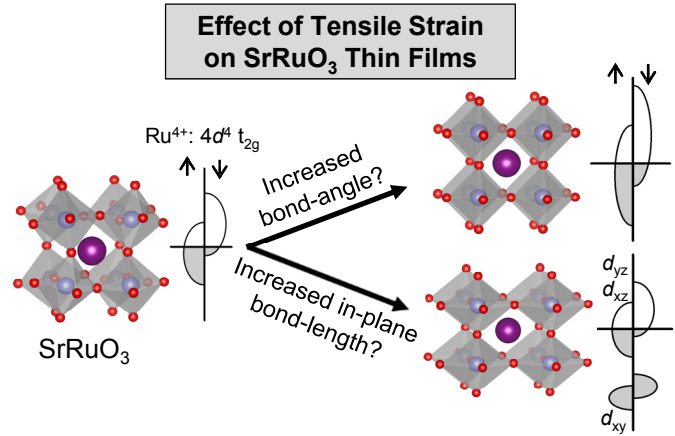

**Fig. S2. Schematic demonstrating the effect of tensile strain on the SRO crystal lattice.** Straining the perovskite crystal structure can induce rotation and tilting of the octahedra and/or elongations of the bond lengths. Here, we show a schematic of the bulk SRO crystal structure and a very simplified picture of two of the possible effects (exaggerated) of applying tensile strain to this structure. One possible outcome is an increase in the in-plane bond angle via octahedral tilts and/or rotations. We can imagine that increasing the bond angle would create a more direct path between sites for conducting electrons to travel, giving an increase in the bandwidth (i.e. hopping integral) and hence a decrease in the resistivity. The other possible outcome is that an elongation of the in-plane bond length can be induced. The bond length elongation in-plane is accompanied by a bond length contraction out-of-plane. The out of plane contraction will lift the degeneracy of the $t_{2g}$ band as the energy of the $d_{xz}$ and $d_{yz}$ orbitals (out-of-plane) will increase as a result of the closer proximity, while the $d_{xy}$ (in-plane) remain unchanged or decrease. This is a well-known phenomenon generally referred to as "orbital selective quantum confinement" (Ref. 21). Therefore, in order to achieve enhanced metallic properties, an increased bond angle is expected. It is important to note that for compressive strain the result is vice versa, i.e. decreased in-plane bond angle or contraction of the in-plane bond length. Therefore, compressive strain is not expected to result in an enhancement of the physical properties. The black arrows indicate the spin of the major/minor carriers.

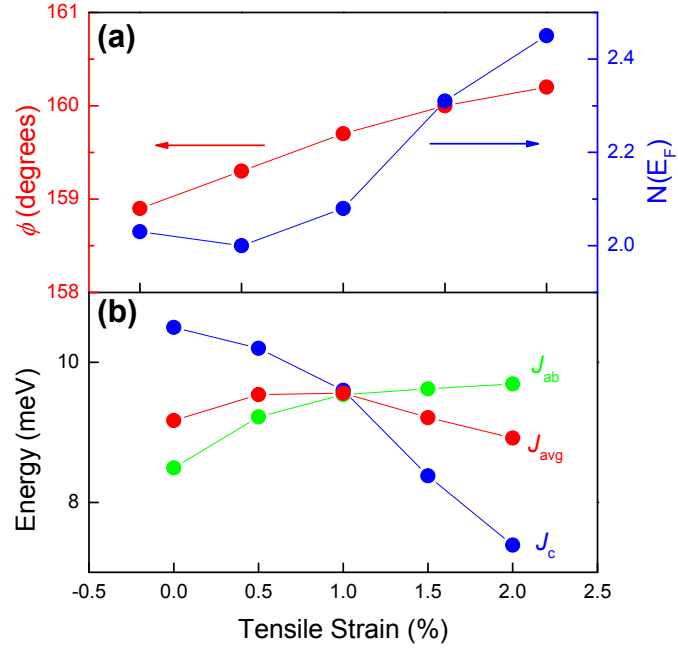

**Fig. S3. First principles calculations of the effect of tensile strain on the in-plane bond angle and exchange energy.** In order to examine the influence of tensile strain on the physical properties of SRO thin films, we performed first-principles calculations as a function of applied tensile strain. The summary of our results are shown here. **(a)** Although there is no appreciable change in the density of states at the Fermi energy, $N(E_F)$, we do notice an increase in the in-plane bond angle which agrees with our predictions. **(b)** Also, note that while the out-of-plane (in-plane) exchange energy, $J_c$ ($J_{ab}$), decreases (increases), the average exchange energy, $J_{avg}$ increases when applying up to 1% tensile strain. SRO is a well-known itinerant ferromagnet and therefore, the ferromagnetic properties should be related to the transport properties, i.e. an increase in conductivity should coincide with an increase in $T_C$, and vice versa. From our calculations we can expect that tensile strained SRO films will follow the later possible scenario from Figure S1 and exhibit enhanced metallic properties. The VASP code (Ref. 24) was used with GGA-PBEsol (Ref. 25) XC functional and the 8x8x6 k-points for structural optimization. The DFT calculations were performed with the range of tensile strain from 0 to 2%

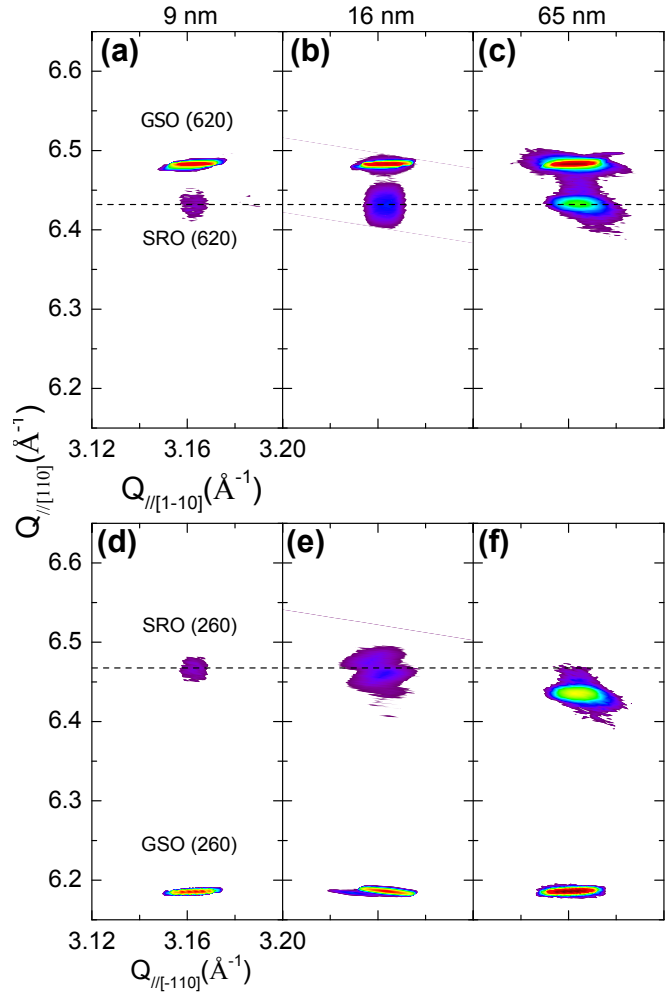

**Fig. S4. Structural phase change as a function of thickness.** Here, we show Reciprocal Space Maps (RSM's) for the $(620)_o$ **(a-c)** and $(260)_o$ **(d-f)** reflections for the 9-, 16- and 65-nm-thick samples. For all three films, the position of the SRO and GSO peaks along the horizontal axis (in-plane) are in the same position, indicating the films are fully strained to the substrate without relaxation. Meanwhile, along the vertical axis (out-of-plane), the 9 nm sample shows that the SRO peaks are located at different positions (Q// = 6.43 and 6.46 nm$^{-1}$ respectively), while the 65 nm film has the SRO peaks in the same location (Q// = 6.43 nm$^{-1}$). In Ref. 18, the authors used RSM and nano-beam electron diffraction measurements and they used this data to show that films thicker (thinner) than 16 nm have higher (lower) crystalline symmetry which is indicative of a structural phase transition to a monoclinic phase as the film thickness is decreased.

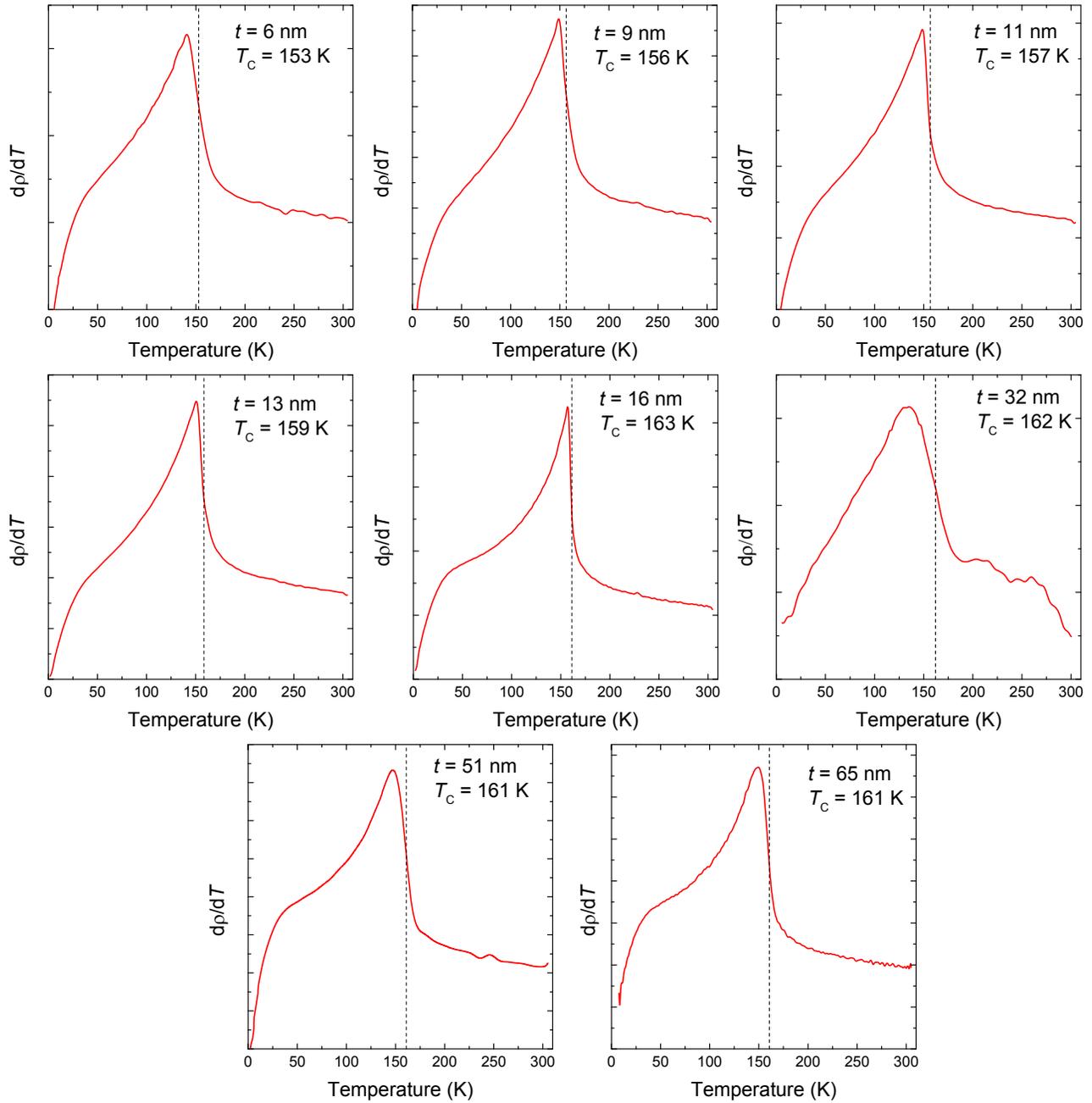

**Fig. S5. First derivative of resistivity vs. Temperature for films grown with optimally moderated deposition rate (~300 pulses per unit cell).** Here we show the method used for estimating the $T_C$ for all of our optimally grown samples. The black dashed lines are guides for the eye.

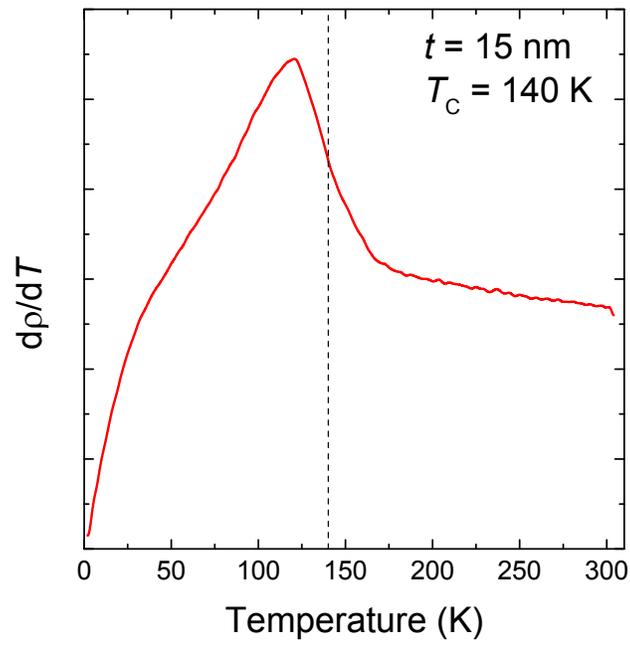

**Fig. S6. First derivative of resistivity vs. Temperature for a film grown with high laser fluence (~3 J/cm²).** In order to demonstrate the effect of laser fluence on the physical properties of SRO epi-thin films we show the estimation of the $T_C$ for this film.

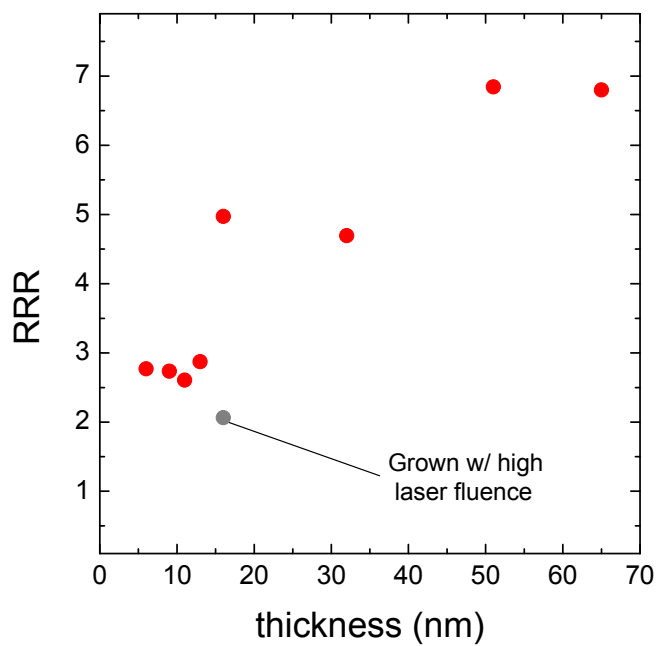

**Fig. S7. Residual resistivity ratio (ρ(300K)/ρ(2K)) (RRR) vs. film thickness.** In general, the RRR can be used as a good indicator of the metallicity of a material. Typical values for "good metals" are 20 and much higher depending on purity and other factors. Previous reports of SRO films and single crystals have observed RRR values from 3-14 (Ref. 11, 18-21) for film and as high as 150 (Ref. 15,16) for single crystals.

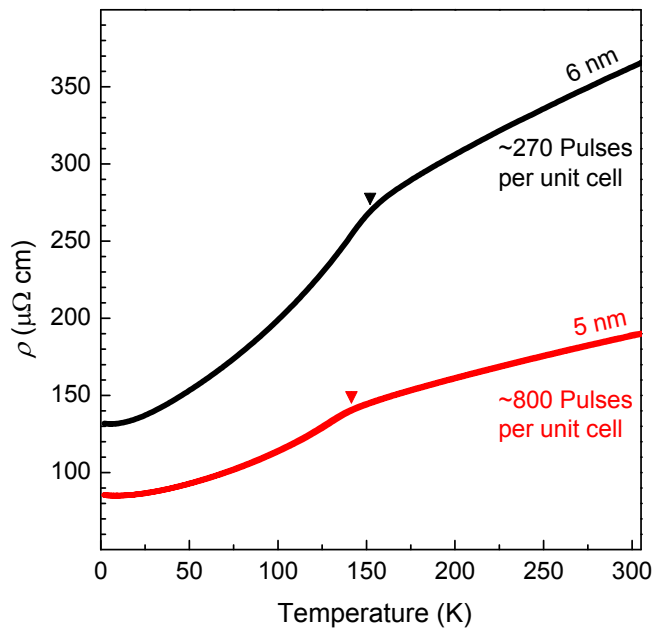

**Fig. S8. Results for ultra-slow deposition rates.** In order to determine the optimal range of deposition rates for SRO epi-thin films, we fabricated many films at different rates. Here, we show a comparison of the $\rho(T)$ data for two films grown at our highest and lowest rates. The black line represents a 6 nm film which was deposited at a rate of ~200 pulses per unit cell, and the red line represents a 5 nm film deposited at a rate of ~800 pulses per unit cell. The films are very similar in thickness and therefore we would expect the resistivity of the two samples to be very similar with the 5 nm film being slightly higher. Surprisingly, we observe that, in fact, the 5 nm film has a much lower resistivity. This is another indicator of the huge impact that the deposition rate can have on the overall quality of the SRO epi-thin films. It is important to note that the slightly reduced $T_C$ of the 5nm film is a result of the thickness limit for ferromagnetism (Ref. 21) in SRO films and not the quality of the film itself.